# Regret Theory And Asset Pricing Anomalies In Incomplete Markets With Dynamic Un-Aggregated Preferences.

Michael C. Nwogugu
Address: Enugu 400007, Enugu State, Nigeria.
Email: Mcn2225@aol.com; mcn2225@gmail.com.
Phone: 234 909 606 8162.

**Abstract[1].**
Although the CML (Capital Market Line), the Intertemporal-CAPM, the CAPM/SML (Security Market Line) and the Intertemporal Arbitrage Pricing Theory (IAPT) are widely used in portfolio management, valuation and capital markets financing; these theories are inaccurate and can adversely affect risk management and portfolio management processes. This article introduces several empirically testable financial theories that provide insights, and can be calibrated to real data and used to solve problems, and contributes to the literature by: i) explaining the conditions under which ICAPM/CAPM, IAPT and CML may be accurate, and why such conditions are not feasible; and explaining why the existence of incomplete markets and dynamic un-aggregated markets render CML, IAPT and ICAPM inaccurate; ii) explaining why the *Consumption-Savings-Investment-Production framework* is insufficient for asset pricing and analysis of changes in risk and asset values; and introducing a "unified" approach to asset pricing that simultaneously considers six factors, and the conditions under which this approach will work; iii) explaining why leisure, taxes and housing are equally as important as consumption and investment in asset pricing; iv) introducing the *Marginal Rate of Intertemporal Joint Substitution* (MRIJS) among Consumption, Taxes, Investment, Leisure, Intangibles and Housing - this model incorporates Regret Theory and captures features of reality that don't fit well into standard asset pricing models, and this framework can support specific or very general finance theories and or very complicated models; v) showing why the *Elasticity of Intertemporal Substitution* (EIS) is inaccurate and is insufficient for asset pricing and analysis of investor preferences.

1. Introduction.
Researchers have since noted the many problems inherent in the one-period capital asset Pricing Model (CAPM) and the one-period Arbitrage Pricing Theory, and have developed the Intertemporal-CAPM, the Intertemporal APT, and the multifactor CAPM. The CML remains a somewhat un-challenged theory that has significant flaws. However, the IAPT, ICAPM and the CML are very inaccurate.

2. Existing Literature.
The existing on the CML, IAPT and ICAPM is extensive, and has been centered on the Consumption-Investment-Production debate. Balvers & Huang (2005) describes the three main classes of research in asset pricing which are as follows. The first line of research is the Consumption-based approach of Breeden (1979) - which is uses a pricing kernel related to the marginal utility of consumption. Initial versions of the consumption-based approach have not worked well empirically, and studies using this approach include Campbell and Cochrane (1999), Lettau and Ludvigson (2001, 2002), and Parker and Julliard (2005). The main problem with this approach is that consumption and its marginal utility are difficult to measure. Balvers & Huang (2005) state that there are asset pricing models that include a production sector but focus predominately on explaining the size of the equity premium. Jermann (1998) argues that both adjustment cost and habit persistence are needed to match the observed equity premium. Boldrin, Christiano, and Fisher (2001) focused on limited inter-sectoral factor mobility and habit persistence in order to explain several aspects of the equity premium and key facts of the overall economy. Jermann (2003) provides an interesting channel through which gains in productivity push up the equity premium given a concave production function, heterogeneous size of the firm in terms of its labor inputs, and a small elasticity of labor. The path breaking model by Gomes, Kogan, and Zhang (2003) provides a structural theory of the risk sensitivities of firms differing by value and size attributes. Balvers & Huang (2005) state that these approaches focusing on the equity premium belong to the consumption-based approach, rather than the production-based approach, because they use the marginal rate of inter-temporal substitution (with or without a habit factor) as the pricing kernel.

---
[1] A later version of this article was published as a chapter in an academic monograph:
Nwogugu, M. (2017). *Regret Theory And Asset Pricing Anomalies In Incomplete Markets With Dynamic Unaggregated Preferences*. Chapter-3 in: Nwogugu, M. (2017). *Anomalies In Net Present Value, Returns And Polynomials; And Regret Theory In Decision-Making* (Palgrave MacMillan).

  The second line of research is the Merton (1973) Marginal Utility approach. However, the marginal utility of wealth is also affected by state variables that indicate how valuable wealth is in different states, but also raises the issue of the importance of state variables - any variables that affect future risk or risk aversion is relevant, and examples include the variables governing habit persistence, irrationality, Regret, utility of investment/savings and the aggregate supply variables affecting productivity. Petkova & Zhang (2005).

  The third line of research is based on production analysis. Liew & Vassalou (2000). Vassalou (2003). Peng and Shawky (1999), Cochrane's (1991). Jermann (2005). Balvers, Cosimano, and McDonald (1990) and Cechetti, Lam, and Mark (1990) argue that aggregate output is equal or proportionate to aggregate consumption and that the marginal utility of consumption can be evaluated at the observed level of output, and thus, aggregate output growth is the main asset pricing factor. The assumed advantage of this approach is that output growth is likely to be measured more accurately than consumption growth. Cochrane (1991,1996) developed a different production-based perspective, which explicitly derives an expression for investment returns and assumes that these can be used to price asset returns. Li, Vassalou and Xing (2003) use a more disaggregated model and found that a four-factor investment-growth approach increases the model fit dramatically. Kim (2003) provides a different theoretical perspective by using duality theory in Cochrane's (1996) framework. Vassalou and Apedjinou (2003) empirically developed a corporate innovation factor and showed that this factor, when added to the market factor, absorbs the momentum effect in cross-sectional asset prices. The Balvers & Huang (2005) model is fundamentally different from the other production-based asset pricing approaches because it derives a pricing kernel based on the state-contingent optimal reactions by firms to productivity shocks; and this kernel is shown to depend largely on factors determining the marginal value of capital. This approach assumes: a) that in a competitive economy with complete financial markets, the marginal rate of intertemporal substitution is tied to a stochastic version of the marginal rate of intertemporal transformation, b) a perfect competition, c) complete markets, d) neoclassical environment where supply shocks in the form of productivity shocks are important.

  Cho (2009) analyzed how Korean households make savings and portfolio decisions (housing plays a special role in the portfolios of households: collateral, a source of service flows, as well as a source of potential capital gains or losses) and also analyzed the role of institutional features by comparing several alternative housing market arrangements to assess their impact on wealth accumulation, portfolio choices, and homeownership. Cho (2009) found that a Lower down-payment also increases both the home ownership ratio and the fraction of aggregate wealth held in housing assets but lowers aggregate net worth with mixed demographic implications.

  Chambers, Garriga & Schlagenhauf (2009a) analyzed how loan structure affects the borrower's selection of a mortgage contract and the aggregate economy; and found that the loan structure is a quantitatively significant factor in a household's housing finance decision; and is dependent on age and income and that these effects are more important when inflation is low. Inflation reduces the real value of the mortgage payment and the changes in the structure of mortgages have implications for risk sharing. Chambers, Garriga & Schlagenhauf (2009b) found that although most countries have tax provisions and subsidies to promote homeownership (which generate an asymmetry in the tax treatment of owner- and rental-occupied housing, which affects the incentives to supply tenant-occupied housing), the progressivity of income taxation can amplify or mitigate the effects of the asymmetries with important implications for housing tenure, housing consumption, portfolio re-allocations, and welfare that differ from those reported in the literature.

The asset pricing model developed in Albuquerque & Wang (2008) focuses on corporate governance issues within a consumption-production-investment framework but**: i)** does not does account for the costs of enforcement of investor protection, **ii)** erroneously assumes that increases in investor protection automatically results in firm growth – differences in investor cognition, investor objectives, ability to process information, and the perceived value of investor protection may dampen positive investor perceptions about firm values, **iii)** over-estimates and over-emphasizes the corporate powers and operational powers of majority shareholders which are quite limited in many common law jurisdictions, and which are also limited even when majority shareholders are managers/officers of the firm (statutorily limited by the "business judgment rule"; fiduciary-duty statutes that apply to officers/directors of corporate entities; and threat of Shareholder derivative lawsuits; and Sarbanes Oxley Act and similar statutes; and investor protection laws enacted by securities exchanges and by securities regulations; and also limited by common-law rules such as the duty of loyalty, fiduciary duties, fraud, usurpation of corporate opportunities; etc.), **iv)** omits the fact that even where there are majority shareholders, third party investors value "independent" boards (which theoretically and actually reduces the influence of majority shareholders), **v)** makes un-realistic assumptions about the portfolio holdings of minority investors and majority investors, **vi)** makes un-realistic assumptions about the path of growth of the firm's dividends and firm value. The utility functions introduced in Albuquerque & Wang (2008) are limited and un-realistic and dont define the scope and ramifications inherent in the agent's decision context.

Miao & Wang (2007) attempted to analyze the joint decisions of business investments, consumption/savings and portfolio selection for an entrepreneur within a real options framework. The asset pricing analysis in Miao & Wang (2007), is flawed because**: i)** as explained in this article, ICAPM/CAPM are incorrect, and thus, the "market portfolio" does not exist in reality; and **ii)** entrepreneurs' inter-temporal wealth allocation/re-allocation decisions are often intertwined with the perceived risks of their ventures and often focus more on the short term (which affects the entrepreneur's utility function and greater-than-normal Regret, both of which were not appropriately defined in Miao & Wang (2007)) – these issues were not separated in Miao & Wang (2007), **iii)** the entrepreneur's utility function and preferences are time-inconsistent because of short-term focus. The utility functions introduced in Miao & Wang (2007) are limited and un-realistic and dont define the scope and ramifications inherent in the agent's decision context. However, Miao & Wang (2007) correctly noted that Real Options theory and approaches erroneously assume that: (**i)** the real investment opportunity is tradable; (**ii)** its payoff can be spanned by existing traded assets; or (**iii)** the agent is risk neutral.

Wang (2009) attempted to analyze an agent's optimal consumption-saving and portfolio choice decisions when he/she cannot fully insure his income shocks and does not know his income growth rate. The relevance of the theories and conclusions in Wang (2009) to real world transactions/events is very limited. Most individuals know or can reasonably estimate any expected growth in their labor income. Most individuals cannot hedge their labor income and dont seek to hedge their labor income. Furthermore, there are very few and very limited insurance markets for hedging of labor income (such markets dont exist in most countries). The utility functions introduced in Wang (2009) are limited and un-realistic and dont define the scope and ramifications inherent in the agent's decision context.

Nelson & Wu (1998) found that the standard intertemporal asset pricing model cannot predict risk premia with the correct sign.

Nwogugu (2006), Nwogugu (2005); Prono (Jan. 2009); Prono (June 2007); Green & Hollifield (1992); Guo (May/June 2004); Kumar & Ziemba (1993); Lewellen & Nagel (2006), Roll (1977); Flam (2010); Gharghori, Chan & Faff (2007); Green & Hollifield (1992); Taleb (2008); Neely, Roy & Whiteman (1999), and Mar, Bird, Casavecchia & Yeung (2009), have shown that the ICAPM/CAPM are inaccurate and thus, the "Market portfolio" is not the most efficient portfolio in terms of risk-reward trade-offs. Some of the issues raised in these articles are also applicable to IAPT, and imply that the IAPT is also inaccurate.

Avramov & Chordia (2006); Fama & French (1996); Gay & Jung (1999); Joyce & Vogel (1970); Moskowitz (2003); Seckin (2001); Brav & Heaton (2002); Brown & Cliff (2004); Hirshleifer (2001); Chen & Cheng-Ho & Jordan (1997); Hong & Stein (2003); Llewellen, Nagel & Shanken (2007); Lettau & Ludvigson (2001); Jaganathan & Wang (2002); Hodrick & Zhang (2001); Ferson, Sarkissian & Simin (2003); Goval & Welch (2008); Campbell & Cochrane (2000); Shanken (1990); Fama (1998); Ehrhardt (Summer 1987); Bray (1994); Campbell, Kazemi & Nanisetty (1999); Braun & Larrain (2005), Iacoviello & Pavan (Nov. 2009), Jamison & Wegener (Nov. 2009), Banerjee (2007), and Donihue & Avramenko (March 2007) documented various anomalies and errors in asset pricing models (mostly consumption-based and production-based asset pricing models). Cochrane (1996) developed an investment based asset pricing model, which performed as well as the CAPM and the Chen, Roll, and Ross (1986) factor model, and performed substantially better than a simple consumption-based model. Attanasio & Paiella (2007) and Gerard & Wu (2006) concluded that Intertemporal Risk is very relevant in Asset Allocation. Detemple & Murthy (1997); Campbell, Kazemi & Nanisetty (1999); Gomes, Yaron & Zhang (2006); Basak & Croitoru (2000); Basak & Croitoru (2000) and Ou-Yang (2005) analyzed the conditions for equilibrium. Battig & Jarrow (1999) introduced a useful definition of market completeness that is relevant to asset pricing. Gomes, Yaron & Zhang (2006) concluded that financing frictions provide an important common factor for the cross section of stock returns and that financial frictions are more important when economic conditions are relatively good.

The CML provides a formulation for the 'optimal portfolio'. The major elements of the CML are:
1) The CML contains some risk free assets.
2) The CML is partly based on the risk free rate.
3) The CML assumes that risky assets earn a return equal to the risk-premium – in this article, the risk-premium will refer to the difference between the return on the market and the risk-free rate ($r_m$-$r_f$).

The ICAPM and IAPT supposedly address the intertemporal nature of decisions and risk management, and provide the 'expected return' for any asset. The ICAPM and the IAPT are supposedly improvements on the CAPM and APT models. The major elements of the ICAPM/IAPT are:
1. The Expected Return is based on the risk free rate. The investor's cost of capital is irrelevant.
2. There are significant "Anchoring Effects".
3. Returns are directly proportional to risk.
4. The Expected Return is largely determined by the returns on the market over multiple periods; or in the case of the APT, by the specified factors over multiple periods.
5. All investors can earn the Risk-Premium – the "risk-premium" will refer to the difference between the return on the market and the risk-free rate ($r_m$-$r_f$).
6. The Beta(s) (ICAPM) and the regression coefficients (IAPT) truly reflects the relationship between the market return (ICAPM) or the factors (IAPT) and the subject asset .
7. There is no utility from hedging.

The SML and the APT provide formulas for the 'expected return' for any asset. The major elements of the SML/APT are:
1. The Expected Return is based on the risk free rate. There are significant "Anchoring Effects".
2. Returns are directly proportional to risk.
2. The Expected Return is largely determined by the returns on the market; or in the case of the APT, by the specified factors. .
3. All investors can earn the Risk-Premium – in this article, the risk-premium will refer to the difference between the return on the market and the risk-free rate ($r_m$-$r_f$).
4. The Beta(s) truly reflects the relationship between the market return and the subject asset.
5. There is no utility from hedging.

3. The Inaccurate Assumptions.
The ICAPM/IAPT and CML are based on many erroneous assumptions, some of which are analyzed as follows – since all of these assumptions are invalid, the CML/ICAPM/IAPT are invalid.

Error Condition-#1: *There Is Continuous Trading And Portfolio Re-Balancing.*
ICAPM, IAPT and the CML can be valid only if there is continuous portfolio re-balancing – ie. portfolio re-balancing is done every one-hundredth of a second or in even shorter intervals. With today's technology, continuous portfolio re-balancing is not possible. Firstly, it takes time to make decisions and process transactions (even electronically processed transactions). Today's trading technology and the duration of back-office processes do not permit such frequent trading. Even if it were valid, the CML, the ICAPM and the IAPT are only a snap shot of the optimal portfolio at a specific point in time. The effective yield on a risk free asset changes over time, and is equal to the stated yield only if the asset is held until "maturity". Similarly, the effective yield on a risky asset changes over time. Hence, for CML, ICAPM and IAPT to be valid, the proportions of the risky and risk-free asset have to be constantly changed, as their yields change over time.

Error Condition-#2: *There Are No Transaction Costs Or Taxes*.
   The ICAPM/IAPT and the CML models erroneously assume that there are no transaction costs. In a world with transaction costs, the expected return, yield on the risk-free asset and the Beta will be affected by transaction costs. The transaction costs will reduce the yields from the risk-free asset, distort the 'risk-premium", and also distort the percentage of total assets invested in the risk free asset and the risky asset. Transaction costs affect the effective sensitivity to the market or to 'factors'.
   The ICAPM/IAPT and the CML formulas/models erroneously assume that there are no income or capital-gains taxes. In a world with either capital gains or income taxes, the effective 'risk-free' rate will be distorted, and the Risk-Premium will be distorted. In a world with either capital gains or income taxes, the effective yields from the 'risk-free' and risky assets will be different from the stated yields; and taxes will distort the proportion of assets invested in both the risky and risk-free assets.

Error Condition-#3: *There Are No Synthetic Securities.*
ICAPM/IAPT and CML models/formulas erroneously assume that there are no synthetic securities. Synthetic securities can be used to replicate a position in the risk-free rate but with higher returns, and the following portfolios are examples – a) a high-yield bond (with a yield of 10-12%) plus a T-Bond futures contract; b) a high yield bond plus a put option or a forward contract; or c) a long term commercial lease plus lease insurance. Similarly, synthetic securities can be used to replicate a long position in the 'market' but with higher returns – ie a combination of index futures and a high-yield stock. With availability of synthetic securities, the purposes/role of the risk free rate and the risk-premium are eliminated.

Error Condition-#4: *There Cannot Be Any Hedging*.
ICAPM/IAPT and the CML can be valid only if there cannot be any hedging. Intertemporal Hedging transforms expected returns by placing a lower bound on returns. With hedging, the risk premium can or will always be positive, and the relationship between the Beta (or in IAPT, any factor) and the expected return is truncated/bounded. Hedging can be costly or costless and this distinction renders IAPT/ICAPM and CML inaccurate. The combination of Hedging and synthetic securities renders the assumptions of ICAPM/IAPT and CML meaningless.

Error Condition-#5: *There Are No Framing Effects.*
ICAPM/IAPT and CML models/formulas can be valid if and only if there are no framing effects.
   In the case of ICAPM, framing pertains to a) the use of the risk-free rate and the market return to calculate the risk premium, b) to the use of the risk free rate, and to the use of the beta. Framing Effects have been defined in the literature.

Error Condition-#6: *Losses Don't Have Any Utility And Don't Cause Regret*.
ICAPM/IAPT and CML models/formulas can be valid if and only if losses don't have any utilities and don't cause Regret. On the contrary, some investors objectives include obtaining losses. With the advent of derivative instruments, losses can be deferred, transferred, and sold. The enforcement of tax laws, causes investors to now focus on after-tax returns and consequences in financial decision making. Hence, in a world in which losses have utility and can cause positive Regret, the concept of "Risk Premium" is moot, because some investors may seek negative Risk Premia, and Beta is meaningless.

Error Condition-#7: *Everybody Can Borrow/Lend At The Risk-Free Rate*.
ICAPM/IAPT and CML models/formulas can be valid if and only if everybody can borrow and or lend at the Risk Free rate. Unfortunately, only a very small percentage of capital markets participants can borrow/lend at the Risk-Free Rate. The Risk Premium is erroneously defined with reference to only the Risk Free Rate, without reference to the investment opportunity set available to each capital market participant, or to the participant's actual cost of capital. There is a finite volume of risk-free assets that can be purchased or sold at any point in time, and so the Risk-Free rate is not available to all capital market investors - this volume of assets changes instantaneously and cannot be controlled by any one capital markets participant.

Error Condition-#8: *All Investors Can Earn The "Risk-Premium".*
ICAPM/IAPT and CML models/formulas can be valid if and only if all participants in the capital markets can earn the "Risk Premium". Firstly the Risk Premium (the difference between the market and risk-free returns) changes instantaneously and cannot be controlled by any one capita markets participant. The Risk Premium changes for each time interval (daily, monthly, quarterly and yearly) and the rate of change over different time intervals is not proportional and is distorted. Not all capital markets participants can earn the Risk premium, because there is only a finite volume of securities/products that can be sold/purchased in order to achieve the Risk Premium. Furthermore, since most indices don't reflect the true nature of the underlying markets, the market return can be achieved only by owning accurate proportions of all securities traded in a market.

Error Condition-#9: *All "Risk-Free" Assets (Typically Treasury Securities) Are Truly Risk-Free; And The Risk Free Rate is Constant.*
      ICAPM/IAPT and CML models/formulas can be valid if and only if all Risk Free assets are truly risk free. "Risk-Free" assets earn the risk-free rate and function as 'risk-free' assets' if and only if they are held until maturity, which usually is not the case in real life. Furthermore, the events that occurred in Italy, Spain and Greece during 2008-2013 show that there are no risk-free assets and that government bonds are risky and default.

Error Condition-#10: Firms Dot Have Any Financing Constraints; And Can Borrow At Any Interest Rate.
ICAPM/IAPT and CML models/formulas can be valid if and only if all firms can borrow any amount of money at any interest rate that they desire – which is inaccurate. Most asset pricing models erroneously assume that firms don't have financing constraints and can borrow at any time, and at any interest rate above the risk free rate. Constraints include – capital markets conditions, legal constraints (loan covenants, statutes, bankruptcy court rules, etc.), credit quality, availability of capital, usury statutes, etc..

Error Condition-#11: *The Risk Free Rate And Beta Remain Constant During All Time Periods.*
      ICAPM/IAPT and CML models/formulas erroneously can be correct if and only if the Risk Free and Beta (or IAPT factors) remain constant over all time periods – which is not possible. Furthermore, many researchers have documented significant problems in calculating Beta  - there are various ways of calculating Beta and there is no industry standard method for calculating Beta. Hence, time-varying Betas are also inaccurate because there is no guarantee that the calculated beat will vary instantaneously with changes in the market  - the models include critical but often false assumptions about the distribution of returns of the "market".

Error Condition-12: *All Portfolios Are Superior To All Individual Assets.*
The CML, IAPT and ICAPM models/formula erroneously assume that all portfolios of assets are superior to holding only one single asset. This property follows from the assume diversification property inherent in the M-V framework. On the contrary, there are assets that may have much better risk-reward profiles than some portfolios.

Error Condition-#13: *All Assets Have The Same "Duration".*
ICAPM, IAPT and CML models/formulas erroneously assume that all assets have the dame duration. The Duration referred to, is the same used in fixed-income analysis. If all Common Stock of all companies don't have the same duration, then assumptions underlying the Beta (and factors in IAPT) and Risk Premium are wrong. Risk Free assets (government bonds and AAA-rated securities) of the same maturity don't have the same duration – the causes of differences in duration include interest payments, call provisions, put rights, form of repayment (cash vs. common stock vs. Pay-in-Kind), etc. Secondly all common stock don't have the same duration - the causes of differences in duration include dividends, corporate by-laws, anti take-over laws, warrants/options issued, trading rules, and perceived risk of the company.

Error Condition-#14: *Interest Rates And The Yield Curve Remain Constant During All Investment Horizons.*
ICAPM, IAPT and CML models/formulas erroneously assume that interest rates and the yield curve remain constant over any time interval. If interest rates change, then there will immediately be arbitrage opportunities between short term and long term securities or across assets.

Error Condition-#15: *All Investors In The Capital Markets Have Constant Amounts of Knowledge.*
CML erroneously assumes that all investors/participants in any given capital market have constant amounts of knowledge and the same amount of knowledge. This is not true because: a) different market participants have different amounts of knowledge and perception of each company/security/asset, b) different market participants have different information processing skills and

Error Condition-#16: *There Are No Limits On The Volume Of Short Positions That An Investor Can Take.*
CML erroneously assumes that there are no .limits on the volume of short positions that an investor can enter into. In reality, investors' possible short positions are limited by: a) availability of securities to short, b) the financing cost of short positions, c) the limitations on available synthetic short positions.

Error Condition-#17: *The Financing Cost Of Short Positions Is Lower Than The Risk Free Rate And The Risk-Premium.*
ICAPM, IAPT and CML models/formulas implicitly erroneously assume that the financing cost of short positions is always lower than the risk free rate and the risk premium. If the converse were true, then there would be arbitrage opportunities – the risk-free rate would be the wrong benchmark for calculating the risk-free premium; and investor could short securities and invest in the risk-free rate.

Error Condition-#18: *Short Positions Are Always Profitable.*
ICAPM and CML models/formulas implicitly and erroneously assume that all short positions are always profitable.

Error Condition-#19: *There Is No Correlation Between The Risk-Free Asset And The Capital Market; And No Correlation Between The Risk Free Asset And The Market Portfolio.*
ICAPM, IAPT and CML models/formulas can be accurate only if and only if the risk free asset in the "market" are not correlated – this is error. Also, researchers have often observed a negative correlation between the short term risk free rates and various proxies for the "Market" such as stock indices.

Error Condition-#20: *For Any Given Time Period, The Rate of Belief-Revision Of Investors Is Higher Than The Rate Of Diffusion Of Information In Capital Markets.*
ICAPM, IAPT and CML models/formulas can be accurate only if and only if the investors' average *Rate of Belief-Revision* ($R_B$) is greater than *Rate Of Diffusion Of Information* ($R_{ID}$) in Capital Markets. On the contrary, research has shown that many investors experience *Inertia* when faced with critical news about their portfolios. *Anchoring effects* and *Framing Effects* also prevent, delay or modify investors' belief revision.
The $R_{ID}$ is the rate at which price-changing information is disseminated among investors, market makers and regulators in the market. Clearly in today's computerized world, information travels very quickly and is almost immediately available to many market participants upon release. Also, various statutes and rules (such as Regulation FD) prevent or discourage insider trading. Hence, in most markets $R_{ID}$ is always greater than $R_B$.

Error Condition-#21: *The Rate Of Substitution Of (A Position In) The "Market" With (A Position In) Any Asset Is Directly Proportional To The Risk Free Rate.*
ICAPM, IAPT and CML models/formulas can be accurate only if and only if the investors' average Rate Of Substitution of (a position in) the "Market" with (a position in) any asset (Ra; expressed as a percentage) is directly proportional to the Risk Free Rate ($R_f$). This is a necessary condition because an investor can hold only a proxy of the Market portfolio (referred to as "Market Portfolio") and if $\partial R_a/\partial R_f < 0$, then i) increases in risk will not be matched by increases in expected return, ii) there will be arbitrage opportunities because arbitrageurs will short the Market portfolio and buy risk free securities when yields of risk free securities are rising, iii) the Market Portfolio will no longer reflect the most optimal portfolio for the average investor; iv) increases in the risk free rate without any asset substitution will increase the investor's margin costs and expected returns and result in mispricing, v) the investors' indifference between holding the market portfolio and holding other assets must be constant or relatively constant in order to derive the cost of capital via ICAPM and IAPT.

Error Condition-#22: *The Rate Of Intertemporal Substitution Of (A Position In) The "Market" With (A Position In) Any Asset Inversely Proportional To "The Risk-Premium".*
ICAPM, IAPT and CML models/formulas can be accurate only if and only if the Rate Of Substitution (replacement) of a position in the "Market" with (a position in) any asset ($R_a$) is inversely proportional to the Risk Premium ($R_p$).
    This is a necessary condition because an investor can hold only a proxy of the Market portfolio (referred to as "Market Portfolio") and if $\partial R_a/\partial R_p < 0$, then: i) increases in risk will not be matched by increases in expected return; and ii) there will be arbitrage opportunities because arbitrageurs will short the Market portfolio and buy combinations of securities and risk free securities that will provide the highest risk premium when yields of risk free securities are rising; and iii) the Market Portfolio will no longer reflect the most optimal portfolio for the average investor. However, this condition can never occur because it implies that the average investor.

Error Condition-#23: *The Average Investor's Rate Of Substitution Of (A Position In) The "Market" With A Position In Any Asset Is Always Greater Than The Average Investor's Rate Of Substitution Of An Equal Position In The "Risk-Free Asset" With (A Position In) Any Asset.*
ICAPM, IAPT and CML models/formulas can be accurate only if and only if the average investor's Rate Of Substitution of a position in the "Market" with a position in any asset ($R_a$; expressed as a percentage) is always greater than the Rate Of Substitution of a position in the "Risk Free Asset" with a position in any asset ($R_r$; expressed as a percentage). This condition ensures and implies that any investor is generally risk averse and is more likely to switch from the Market Portfolio to an asset A, than from a risk free asset to A in any market. Hence, the indifference curve of the risk free asset and the Market Portfolio must be downward sloping.

Error Condition-#24: *The Investor's Actual And Marginal Borrowing Rate Is Irrelevant To His/Her Expected Return From Any Asset/Portfolio.*
ICAPM, IAPT and CML models/formulas can be accurate if and only if the average investor's marginal borrowing rate is irrelevant to his/her expected return. ICAPM and CML don't incorporate investors' Marginal Borrowing Costs. Most IAPT models also don't incorporate the average investor's Marginal Borrowing costs and changes in such costs over time. This condition cannot be feasible because, any rational or irrational investor will typically consider his/her marginal borrowing costs (including margin costs in securities accounts) - such investments may require additional capital and thus borrowing (as in futures accounts or real estate development projects or acquisitions of assets).

Error Condition-#25: *Investors And Traders Don't Experience Regret; And Regret Does Not Affect Investor's Expected Returns.*
ICAPM, IAPT and CML models/formulas can be accurate if and only if the average investor or trader does not experience Regret, and Regret does not affect their expected returns. ICAPM, IAPT and CML completely omit Regret. There is a substantial literature on the effects of Regret on human decision making – extendible to consumption choices and portfolio selection/re-balancing decisions. Hence, this condition is not feasible.

Error Condition-#26: *For Any Asset, The Average Investor's Marginal Propensity-To-Substitute (For Any Other Asset) Is Irrelevant To Calculation Of Its Expected Return.*
On the contrary, the average investor's Marginal propensity-to-substitute any asset is highly relevant to both the investor's horizon, opportunity costs, indifference to other assets/opportunities and thus, his/her expected returns. Hence, ICAPM, IAPT and CML models/formulas are very inaccurate

Error Condition-#27: *Expected Returns Are Only In The Form Of Cash; Investors Don't Experience Any*

*Positive Utility From Holding Assets.*
On the contrary, it has been documented in the literature that investors can and do gain non-monetary utility from holding specific assets and specific combinations of assets. Such non-monetary utility is also part of investor's return from holding the asset, but is not incorporated into ICAPM, CML and most IAPT models.

Error Condition-#28: *There Is No Or Minimal Correlation Between The Asset-Beta And The Risk Free Rate.*
On the contrary, the Betas of some companies are highly correlated with the risk-free rate and this renders the ICAPM and CML inaccurate. Such companies include finance companies, banks, mortgage REITs, etc.. If the Beta is correlated with the Risk-Free rate, then the Risk Premium will be distorted and inaccurate, and some of the terms in the ICAPM and CML formula will also be distorted.

Error Condition-#29: *In The IAPT, The Nature Of The Relationships Indicated By The Regression Coefficients Remains Constant Over Time; And For All Assets, There Is Minimal Multi-Collinearity Among Factors.*
On the contrary, the relationships indicated by the IAPT regression co-efficients are very dynamic and change constantly – some change continuously. Also, for any asset, there is often substantial multi-collinearity among the regression factors. Hence, IAPT is inaccurate.

Error Condition-#30: *In The IAPT, The First Derivative Of The Number Of Factors With Respect To The $R^2$ Of The Equation, Is Constant.*
The IAPT models erroneously assume that the first derivative of the number of factors with the $R^2$ of the IAPT regression equation remains constant over time – ie. that $\partial R^2/\partial n = 0$, where n is the number of factors.

Error Condition-#31: *The Risk-Free Rate Always Adequately Incorporates Inflation Risk And Horizon Risk; Or* There Is No Inflation Or Deflation
On the contrary, the Risk-Free Rate does not always incorporate inflation risk and horizon risk – ie. the inflation protected US Treasury securities. Given such omission, the Risk Premium will always be inaccurate and the ICAPM and CML are inaccurate. Most asset pricing models don't incorporate the effects of inflation, which in emerging economies can range from 10% to 3,000% annually. Inflation/deflation: a) affects investor expectations of asset returns, and asset prices, b) inflation reduces investors' real returns, c) affects the possibility and availability of an efficient hedge for investments, d) affects investor current and future consumption, demand for goods and services, output, interest rates and availability of capital. Hence, most asset pricing models are grossly mis-specified.

Error Condition-#32: *The Investor's Investment Horizon Does Not Matter; And The Changes In the Investor's Preferences And Risk Tolerance Do Not Matter; Intertemporal Risk And Benefits Can Be Defined Solely In Terms Of Standard Deviation, Mean Return, And Consumption*
Investors' investment horizons matter and affect their expected returns because there are opportunity costs, Regret and utility/disutility from holding assets. ICAPM, IAPT and CML are defined only with respect to specific horizons, and don't account for changes in investor horizons. In most asset pricing models, the indicator of consumer/investor state and Preferences is Utility, which is expressed primarily in terms of standard deviation, means and consumption. This approach is incorrect. Investors' "preferences" and "States" can also be expressed with other metrics such as Regret, Opportunity Costs, downside Risk, Willingness-To-Accept-Losses, etc..

Error Condition-#33: *All Markets Are Efficient In All Consecutive Periods; And The* "No-Arbitrage" Condition Exists In Consecutive Time Intervals
There has been significant research that has proved that markets can be, and are inefficient. The main problem is that research on market efficiency has always erroneously assumed constant knowledge and unity of opinions among market participants and regulators; and that all existing market inefficiencies are instantly recognized and taken advantage of by market participants. In reality, there are: a) significant differences in knowledge of various classes of market participants – individual investors, traders, etc., b) not all market inefficiencies are identified and taken advantage of - due to knowledge limitations, availability of capital, time, regulations, Regret, Risk Aversion, inaccurate computer models, etc., c) the sheer volume and instantaneous changes in psychological states of investors and continuous revisions of investor beliefs, creates significant differences in opinion and hence, arbitrage opportunities.

Error Condition-#34: There is Equilibrium In Financial Markets.
Most asset pricing models erroneously assume some degree of "equilibrium" in the economy and in some markets. Contrary to generally accepted assumptions, there cannot be any "equilibrium" in financial markets because of the following reasons: a) there are instantaneous changes in investor perceptions, aspirations and beliefs about futures states, such that even dynamic equilibrium cannot exist; b) the definitions of demand and supply are often based partly or wholly on the amount of capital in the market, but does not include the amount of capital that is potentially available to all market participants, c) the existence of derivatives eliminates the possibility of equilibrium, d) the possibility of shorting securities and the interest charge for short positions eliminate the possibility of equilibrium, e) inflation (particularly in emerging market economies) has a significant but un-recognized effects on asset prices and asset values.

Given that none of the above mentioned conditions are feasible, all or most existing asset pricing models are inaccurate. This has substantial implications for asset management, dividend policy, capital budgeting and risk management.

4. The Consumption-Savings-Investment-Production Dichotomy Is Inaccurate.
During the last one hundred years, Economists and central bankers have built most of their models and analysis on the Consumption-Savings-Investment-production (the "CSIP") dichotomy which treats each of the four factors as an almost unique "domain" of analysis (and does not focus on the inter-connectedness of the four factors); and which includes supply-side and demand-side analysis. However, the following economic catastrophes have proven that the CSIP is inefficient and flawed in many ways:

    i) The Asian financial crisis of the 1990s.
    ii) The collapse of the LDC debt market in the mid-1990s.
    iii) The US recession of the early 1990s.
    iv) The stagnation of the growth of the Asian Tigers.
    v) The lack of improvement of the quality of life in most large developing countries (India, China, Brazil, and Indonesia) despite reported growth in their GDP/GNP.
    vi) Significant Trade Deficits in many countries during the last ten years.
    vii) The sub-prime mortgage crisis in the US during 2005-2010; and the collapse of asset securitization markets in the US during 2008-2009; and the insolvency of Fannie Mae and Freddie Mac in the US.
    viii) The global financial crisis that began in 2007.
    ix) The failure of government stimulus programs in the EU, US, Japan and other countries during 2008-2012.
    x) The economic problems in Greece, Ireland, Iceland and Spain during 2008-2012 (bailout of Greece by the EU; and downgrading of Spain's debt to below-investment grade).
    xi) The economic problems in Latin American countries during the 1990s.
    xii) Significant increases in amounts of government debt in many countries during the last ten years.
    xiii) Un-recorded inflation in many countries; and continuing hyper-inflation in many developing countries.
    xiv) Under-developed or non-existent real estate markets in many countries.
    xv) High unemployment in many countries during the last ten years.
    xvi) The inaccuracy of the major rating agencies which contributed to many financial failures, and to market participants' inability to properly assess risk.
    xvii) The adverse effects of the fixed exchange rate of the Chinese currency (Yuan).
    xviii) The loan losses incurred by Japanese banks during the 1990s.
    xix) The "informal black-market" economies in many developing countries.
    xx) The last-resort "dollarization" of the economies of many developing countries.
    xxi) The Russian financial crisis of 1998.
    xxii) The Nigerian financial crisis of 2005-2012.

xxiii) The crash of the US technology stock market in 2000.
xxiv) The crash of Chinese stock markets during 2015.
xxv) The sovereign debt crisis of Greece, Italy and Spain during 2010-2015.

The CSIP dichotomy has always been used, and Consumption has always been analyzed from the perspective of the household without detailed analysis of the nuances of decision making and the psychological benefits/costs of the allocation of wealth. Several studies that have analyzed the differences between consumption (of different goods/services) and investment conclusively show that consumption and investment/Savings patterns are not uniform across goods/services, financial products, time, industries and location; and that income and savings patterns are not uniform across time, region, industry, and age, both in terms of the actual dollar amounts and the rationale for such behaviors and the utility/disutility derived from such behaviors. Hence, the terms "aggregate consumption" and "aggregate Investment" and "aggregate income are misnomers within the context of asset pricing.

The Consumption-Savings-Investment-Production (CSIP) framework and dichotomy/debate may have been useful in past eras where the following conditions existed:
1) Information was limited and information diffusion was much slower – there was no internet, and it was difficult to obtain statistics about markets and products/services, and to compare.
2) There was more uniformity of products and services – today modern technology enables companies to provide a much wider variety of goods and services at lower per-unit costs.
3) Payment systems were limited – today there is a proliferation of payment systems
4) Entertainment was limited – today, the form, access and pricing of entertainment are drastically different and there has been an exponential proliferation of the types and volume of entertainment.
5) For any given product, the sources of utility/disutility were limited.
6) Loan volumes were much smaller – today, loan volumes in various countries are much larger.
7) There were fewer complex financial instruments.
8) Stable taxation – taxation was simpler. Today, taxation (personal, business, income, capital gains, etc.) has become very complex.
9) There were fewer ways to hedge financial risk and operational risk.
10) The traditional central banking tools of monetary policy were more effective.
11) Government deficits were generally smaller.

4.1. Savings.

Savings is much less of an important economic indicator for several reasons. First, there are now a wider variety of sources of capital for investment and lending. Companies, foundations, pension funds, insurance companies and governments (local, state and federal) now have active treasury functions and routinely invest in all types of securities across different maturities and countries, and have essentially replaced household savings as the primary source of capital. Secondly, there has been a proliferation of new forms of financial products (such as 100% LTV loans; and long-term leasing) that don't require any equity investment by households or companies. These loans have substantially reduced the incentive to save and the importance of savings. Thirdly, the prevalence and rapid growth of online and non-internet social networks has reduced the "need" for household savings –if people can borrow short-term loans from friends and obtain other goods (temporary housing, use of vehicles, etc.) from friends, then the need to save for emergencies and contingencies declines. Fourth, the growth of finance companies that provide short term loans to individuals has also reduced the need for, and relevance of savings as an economic indicator.

Fifth, there has been an increasing greater divergence between savings and consumption during the last twenty years - savings in the traditional sense does not equate to changes in consumption patterns in the future – rich and super-rich households still borrow for various purposes. Hence the relationship between savings and consumption or investment has become much more tenuous than in the past due to changes in the financial services sector, and the utility of debt/borrowing.

Sixth, the utility of savings has declined during the last twenty years. In the past, savings provided some assurance of a safe retirement and ability to manage contingencies and pay for household necessities such as education and healthcare costs. However, un-recorded excess inflation in developed and developing countries (which has been non-uniform across industries and goods), higher taxes, financial contagion, volatility in capital markets and declining financial stability of banks has reduced the utility derived from savings.

      Seventh, the increase in government financial support for senior citizens particularly in developed countries has reduced the incentive to save and the utility of savings. Most of this government support is funded by increased taxes. Eighth, the proliferation of insurance and reinsurance products (for long term care, senior citizen's housing, assisted living, etc..) has reduced the incentive to save money and the utility of savings. Ninth, "lending" by households (via the purchase of bonds and notes) is often not perceived as a form of savings (both by households and central banks that calculate national accounts). There is increasing blurring of the differences between "household Investment" and "household savings". Tenth, the perceived utility of a securities brokerage account is very different form the perceived utility of a traditional savings account at a bank (and from that of home equity). Given that the trend in consumer financial services has been a major shift away from savings accounts to brokerage accounts, the utility gained from savings has clearly declined.

The transferability of the utility of savings has declined during the last twenty years primarily because of:
 a) Substantial divergences among individuals about the value and utility of traditional savings.
 b) Substantial divergences among individuals about the utility of traditional measures of wealth (cash, securities, home equity, intangibles, savings accounts, gold/silver, etc.).

The utility of savings is derived as follows.

Let:
$U_s$ = utility of household Savings
$I_s$ = PV of savings.
$X_e$ = PV of expected future expenses not covered by savings.
$X_u$ = PV of unexpected expenses.
$X_i$ = PV of inflation effects.
$L_f$ = Negative effects of general instability of financial system on savings and home equity – similar to Regret.
$V_h$ = Present value of Expected home equity.
$t$ = time to death (years)
$r$ = discount rate
$I_g$ = PV of expected government support for senior citizens.
$I_i$ = PV of third-party insurance benefits.
$I_s$ = availability of short term un-secured loans in the future

$U_s = \exp[\int_0^t (I_s - X_e - X_u - X_i - L_f + V_h - I_g - I_i)\, dt]$

This implies that most of the utility functions derived for savings and investment in the existing literature are inaccurate and insufficient to describe the dynamics of real world conditions.

4.2. Aggregate Investment And Investment.
Aggregate Investment is also a much less useful economic indicator for many reasons. Aggregate Investment lumps industry investment together with consumer investment. Aggregate Investment does not distinguish between differences in investment objectives and the structure of the investment (which is increasingly critical point of differentiation). Secondly, Aggregate Investment is not adjusted for the duration of the investment, which is a critical element of the economy. Third, Aggregate Investment does not reflect government subsidies and incentives for both households and companies – this omission distorts true nature of economic activity. Fourth, Aggregate Investment does not distinguish among capital investment and investments in securities, and investment in intangibles.

4.3. Intangibles - The Production And Consumption Of Intangibles Differs From General Consumption And Traditional Production And Investment.

Wyatt (2005), and Wong & Wong (2001) found significant economic and behavioral effects from accounting recognition/non-recognition of intangibles. According to Salinas (2009) (and other studies), Intangible Assets constitute 60% to 75% of the market capitalization value of the major stock indices in the World; and thus changes in the disclosed values of Intangible assets can affect individual and group psychology. Corrado, Hulten & Sichel (2011), and Slaughter (2013) found that Intangible contribute substantially to the US economy. At the individual level, Intangibles also include Social Capital and personality traits that account for success or failure in business transactions and personal relationships. Unlike traditional goods, the consumption of Intangibles occurs over many periods, and can yield future utility in the form of: a) home equity, b) social networks, c) social capital and Reputation; d) peace of mind; e) reduced Regret; f) skills which improve labor mobility; g) second income; etc.

The production of intangibles differs from traditional production. Substantial and increasing percentages of Intangibles are produced in the services sector, or through services (non-manufacturing) activities such as software development, advertising, promotions and social media networks. The "consumption" of Intangibles also differs from traditional consumption in terms of timing, place and frequency. In many instances, Intangibles production is involuntary or incidental/tangential. The Intangibles production decision is debated within the household and companies, and is sometimes a major component of the identity and self-worth of such units. Hence, there are more psycho-social processes (internal and external) associated with Intangibles consumption; than is indicated in the existing literature.

Similarly, the Intangibles consumption decision is debated within the household, and is a major component of the identity and self-worth of the household. Hence, there are more psycho-social processes (internal and external) associated with Intangibles consumption; than is indicated in the existing literature.

Unlike traditional goods and traditional Savings, Intangibles consumption has substantial, sometimes irreversible and very observable effects on the social networks of households. Furthermore, most analysis of Consumption and savings don't account for the fact that behavioral factors such as shocks (loss of income; ill health, property damage), and changes in tastes/preferences of households, etc., are critical determinants of Intangibles consumption. Unlike traditional goods and traditional Savings, Intangibles consumption has substantial, sometimes irreversible and very observable effects on the social networks of households.

De Roon & Szymanowska (2012) found that when U.S. stock portfolios are sorted according to size; momentum; transaction costs; market-to-book, investment-to-assets, and return-on-assets (ROA) ratios; and industry classification, the portfolios show considerable levels and variation of return predictability, that is inconsistent with asset pricing models, such that the risk premium predicted by asset pricing models is not sufficient compensation for systematic risk. In addition to short sales constraints, holding periods and transaction costs, the asset pricing anomalies stated in De Roon & Szymanowska (2012), Hodrick & Zhang (2001), Maslov & Rytchkov (_____), and Lewellen & Nagel (2006) can also be explained by differences in perceptions of, and amounts of Intangibles (ie. valuation, volatility; risk).

4.4. Leisure Differs From General Consumption, Investment And Production.

The substantial differences between the consumption of Leisure and traditional consumption, have not been properly addressed in the existing literature on asset pricing. Unlike traditional goods, the consumption of Leisure occurs over many periods, and can yield future utility in the form of: a) social networks, b) social capital, c) peace of mind, ) Reputation; e) reduced Regret; f) second income – from additional skills; etc. In many instances, the Leisure consumption decision is debated within the household, and is a major component of the identity and self-worth of the household. Hence, there are more psycho-social processes (internal and external) associated with Leisure consumption; than is indicated in the existing literature. The nature and timing of Leisure activities have changed substantially since the mid 1990s and the advent of Broadband Internet. Information about more Leisure opportunities are now available on the Internet, which also provides various platforms for matching individuals/households with low-cost leisure opportunities. Movies, Games, group memberships and social networks are readily available on the Internet. These trends have generally resulted in segmentation of Leisure, and declining costs of Leisure for many classes/types of Leisure activities.

Also, more Leisure activities are or can become income producing activities or "home production" – this is partly because: i) issues like marketing, advertising, distribution, customer services and quality have all been made cheaper and more available to small businesses by the growth of the Internet; ii) un-even growth in income and living expenses (across industries, regions; households) and shocks (such as the subprime crisis) have compelled adults to seek second incomes. Aguiar & Hurst (2007). Gelber & Mitchell (2012). Gronau (1977). Unlike traditional goods and traditional Savings, Leisure consumption has substantial, sometimes irreversible and very observable effects on the social networks and future allocations and consumption choices of households. Furthermore, most analysis of Consumption, Investment and savings dont account for the fact that behavioral factors such as shocks (loss of income; ill health, property damage), and changes in tastes/preferences of households, etc., are critical determinants of Leisure consumption. Unlike traditional goods and traditional Savings, Leisure consumption has substantial, sometimes irreversible and very observable effects on the social networks of households.

In addition to short sales constraints and transaction costs, the asset pricing anomalies stated in De Roon & Szymanowska (2012), Hodrick & Zhang (2001), Maslov & Rytchkov (_____), and Lewellen & Nagel (2006) can also be explained by differences in perceptions of, and amounts of Leisure at the individual, household and company levels (ie. value of Leisure; gains/losses from Leisure activities; risk; etc.).

### 4.5. The Consumption Of Housing Differs From General Consumption And Traditional Savings.

Another critical issue is that the substantial differences between the consumption of housing and traditional consumption, have not been properly addressed in the existing literature on asset pricing. Housing accounts for 25%-40% of the economies of many developed countries and third world countries. Furthermore, housing accounts for more than half of the total Wealth of most households in most developed countries, and is the biggest investment decision made by many households. El-Attar & Poschke (2011). Housing-related monthly expenditures account for 20%-40% of the total monthly household expenditures in most countries. The concept of "Housing" can also be extended to companies, for which occupancy costs (rent; maintainance; overages; utilities; property taxes; fees; etc.) account for 15%-50% of monthly operating expenses. Yang (2009) contrasted the consumption of housing and that of traditional goods. In addition to the differences noted by Yang (2009), the following are other critical differences among the consumption of housing and the consumption of traditional goods, and traditional Savings.

Unlike traditional goods, the Housing unit is typically fixed in time, space and form. The Buyer usually cannot change the configuration of, or move the housing unit. The transaction costs and search costs for buying/selling housing units are relatively large, and can vary drastically across geographical regions and time. The purchase of a housing unit almost always involves a mortgage loan (which incurs additional search costs, processing costs, reputation costs, household dynamics, and commitment costs). In most jurisdictions, the purchase of a housing unit incurs future property taxes. The consumption of housing is highly regulated at the local, state and federal levels – by zoning laws, building codes, environmental laws, ordinances, mortgage laws, etc..

Unlike traditional goods, the purchase of housing typically involves as much "consumption" as the sale of the same housing unit. This is henceforth referred to as "*asymmetrical two-sided housing transaction consumption*".

Unlike traditional goods, the consumption of housing occurs over many periods, and can yield future utility in the form of: a) home equity, b) social networks, c) social capital, d) peace of mind, e) reduced Regret, etc. In many instances, the housing consumption decision is debated within the household, and is a major component of the identity and self-worth of the household. Hence, there are more psycho-social processes (internal and external) associated with housing consumption; than is indicated in the existing literature.

Unlike traditional goods and traditional Savings, Housing consumption has substantial, sometimes irreversible and very observable effects on the social networks of households.

Furthermore, most analysis of Consumption and savings dont account for the fact that behavioral factors such as shocks (loss of income; ill health, property damage), and changes in tastes/preferences of households, etc., are critical determinants of housing consumption. Traditional analysis of Consumption and Savings dont differentiate among housing as physical space/shelter, housing as "expectations" and investment; housing as "conformance/status", and housing as a bundle of psychological/social goods. Traditional analysis of Consumption and Savings dont analyze the various effects of debt and access to credit on the consumption of housing and non-housing goods. All else held constant, the consumption of housing varies drastically among different types of housing (condos vs. coops. vs. townhouses vs. rental units vs. single family homes).

Housing remains a basic element of human needs and economic activity but its ramifications and somewhat unique results are not fully understood. There remains several paradoxes in the housing markets that transcend location, time, wealth and household structure such as the following: **a)** why households that rent housing units are not taxed but households that own homes are taxed without any consideration of how the household financed the home purchase, and regardless of whether or not home-ownership is less or more beneficial for the economy; **b)** for some wealthy households, the stock of housing is potentially "infinite" if its assumed that exchange costs (costs to change housing units) are relatively minimal, and the utility of exchanges is substantial and transferable; **c)** rent controls are not enforced fully such that even middle income households benefit from rent control, **d)** rent control/stabilization laws are almost "permanent" in most large cities in developed countries, and many of the specific rent control mechanisms are not designed to vary significantly with economic cycles or time, **e)** in most countries, the federal/central governments continue to delegate construction/maintenance of housing almost entirely to the private sector even though its apparent that private companies don't have sufficient incentives to provide truly affordable housing, **f)** federal and state governments give out tax credits to developer to build affordable housing, but don't give tax credits to house holds to limit or optimize their consumption of housing (there is huge "housing waste" caused by empty nesters and mismatch of housing needs and housing supply), g) in many countries, despite the critical nature of housing, governments are hesitant to actively participate directly in the housing sector, and government activity in the housing sector is limited to income support (housing vouchers, free emergency housing; tax credits) and offering free land, and to a lesser extent, low-cost financing); h) with all other factors assumed to be constant and similar across housing types (including price), the consumption of housing varies drastically among different types of housing (condos vs. coops. vs. townhouses vs. rental units vs. single family homes

In addition to short-sales constraints and transaction costs, the asset pricing anomalies stated in De Roon & Szymanowska (2012), Hodrick & Zhang (2001), Maslov & Rytchkov (_____), and Lewellen & Nagel (2006) can also be explained by differences in perceptions of, and amounts of Housing at the individual and household levels (ie. value of Housing; gains/losses from Housing; risk; etc.); and the amount of occupancy costs for commercial real estate at the company level.

5. Regret Theory, Prospect Theory And Behavior-Based Asset Pricing Models.

Given the many problems inherent in existing asset-pricing Models, Regret Theory is a viable alternative to decision model. This is because Regret Theory incorporates many of the behavioral/psychological issues that ICAPM and IAPT Models do not or cannot capture – such as flexibility, Real Options, Regret minimization; reversibility of decisions, dynamic cost-of-capital, Framing effects, etc.. This section surveys some critical literature on Regret Theory to show how it has been used in asset pricing and for project selection/evaluation.

Bonomo, Garcia, Meddahi & Tédongap (2011); Hirshleifer (2001); Ray & Robson (2012); Jamison & Wegener (Nov. 2009); Han & Yang (2013); Barberis, Huang & Santos (2001); Korniotis & Kumar (2011); El-Attar & Poschke (2011); Berkelaar & Kouwenberg (2009); Solnik & Zuo (2012), have analyzed the effect of individual and group behaviors on asset pricing. Solnik & Zuo (2012), Bonomo, Garcia, Meddahi & Tédongap (2011); Lia & Yang (2013); Dodonova & Khoroshilov (______); Barberis & Huang (2008); Hung & Wang (2005); De Giorgi, Hens & Mayer (2007); Yogo (2008); and Barberis, Huang & Santos (2001), developed behavior-based asset pricing models, which were inspired by Regret Theory or Prospect Theory.

Nasiry & Popescu (2009) analyzed the effects of anticipated regret on consumer decisions, firm profits and policies, in an advance selling context where buyers have uncertain valuations; and found that advance purchases trigger action regret if valuations are lower than the price paid, and delaying purchase causes inaction regret. Nasiry & Popescu (2009) developed a Regret threshold above which firms should only spot-sell to homogeneous markets, and otherwise advance selling is optimal. Nasiry & Popescu (2009) also found that the effect of regret on profits depends on the type of regret, market structure and the firm's pricing power - and Action regret lowers the optimal profits of a price-setting firm in homogeneous markets, while inaction regret has the opposite effect. According to Nasiry & Popescu (2009), firms can benefit from regret by creating a buying frenzy, where consumers purchase in advance at negative surplus; and Action regret can be profitable if high valuation consumers are more regretful, or if the firm is price-constrained.

Michenaud & Solnik (2008) applied Regret Theory to derive closed-form solutions to optimal currency hedging choices – they theorized that Investors experience regret of not having chosen the ex-post optimal hedging decision; and thus, investors anticipate their future experience of regret and incorporate it in their objective function. Michenaud & Solnik (2008) derived a model of financial decision-making with two components of risk: traditional risk (volatility) and regret risk; and their results contradicted traditional expected utility, loss aversion, and Dis-appointment-Aversion theories.

Ghosh (1993) used Regret Theory to explain some survey results on managers' dividend policy decisions under uncertainty; and postulates that (1) the decision to pay dividends and simultaneously raise venture capital from external sources is attributable to managerial aversion for regret at the failure of a risky investment opportunity implemented with internal funds generated by a conservative dividend policy; and (2) the decision to support dividends with borrowed funds when earnings are declining is motivated by the prospect that an improvement in the firm's financial condition will make the managers proud that their judgment has helped avert a potential crisis for the firm without any wealth loss to its shareholders.

Muermann, Mitchell & Volkman (2006) showed that anticipated disutility from regret can have a significant and drastic effect on investment choices. They analyzed and modeled how plan-participants' asset allocation decisions in a defined contribution (DC) pension plan might vary with their preferences about risk and Regret. Muermann, Mitchell & Volkman (2006) found that the Regret-Averse investor will typically hold more stock than a risk-averse investor when the equity premium is low but less stock when the equity premium is high; and that Regret increases the regret-averse investor's willingness-to-pay for a guarantee when the portfolio is relatively risky, but decreases it when the portfolio is relatively safe.

Brocas & Carrillo (2005) analyzed decision making by a hyperbolic discounting agent; and showed that the agent may rationally decide to consume with negative expected NPV only to prevent himself from consuming in the future which could be profitable from a future perspective but highly detrimental from the current viewpoint. Comparative statics reveal that the value of information is U-shaped.

Nwogugu (2006) introduced new models of Regret and Willingness To Accept Losses (WTAL). The Nwogugu (2006) Regret model can be used for project evaluation/selection.

Dodonova (2009), DeKay (2009), Bleichrodt, Cillo & Diecidue (2010), all analyzed the use of Regret Theory for project evaluation and project selection. Chandrasekhar, Capra, Moore, Noussair & Berns (2008), and Canessa, Alemanno, Motterlini, et. al. (2009), and Coricelli, Dolan & Sirigu (2007) analyzed neuro-biological Regret.

Gollier & Salanié (2006); Sagi & Friedland (2007); Wang, Triantaphyllou & Kujawski (2008); Huang, Tzeng & Liu (2009); and Laciana & Weber (2008) analyzed Regret. Frehen, Hoevenaars, Palm & Schotman (2008) analyzed regret aversion in the retirement investment decision of defined contribution plan participants; and developed and priced a look-back option on a life annuity contract.

6. Definitions.
*Definition-1*: The Consumption-Savings-Investment-Production Dichotomy (the "CSIP") is irrelevant because all four asset-pricing approaches are elements of one *Unified Intertemporal Wealth-Allocation Decision* ("UIWD") by an investor or household. UIWD encompasses all the elements of the household's or the firm's simultaneous decisions about Consumption, Savings, Investment, Taxation, Production, Leisure, Intangibles and Housing. The investor or head of the household typically budgets for each period (weekly or monthly or annually) either mentally, or in physical form (written budgets). Such budgets are often discussed with others – including spouses/partners, investment advisors or other members of the household. The investor/household-head decides how much of his/her wealth to allocate to Consumption, Savings, Investment, Production, Leisure, Intangibles and Housing. Such decisions may be revised/updated during the subject period, and the initial allocation decision is followed almost immediately by more detailed sub-allocations within each of the six domains (*Consumption* – household necessities, luxuries, etc.; *Investment* - production, amount of work time, over-time, bonuses, savings, amounts to allocate to or withdraw from various savings programs, tax considerations, etc.; *Taxes* – capital gains taxes, income taxes, sales taxes, property taxes, etc.; *Leisure* – entertainment, vacation, etc.; *Intangibles* – training, conferences, professional certifications, cost of patents/trademarks, etc.; *Housing* – rent, mortgage payments, maintenance, etc.).

*Defition-2*: Total Wealth (W(.)) includes monetary and non-monetary wealth (time; Intangibles; Intellectual Property Rights; social capital; contingent rights; utility from deferral of obligations; etc.). Investable Wealth includes wealth that can be readily converted into cash or can be readily exchanged for other assets.

*Definition-3*: Housing for the agent refers to all costs necessary for tenancy or ownership of a housing unit (individuals) or office space (companies) – such as rent, property taxes, mortgages payments, home insurance, maintenance costs, etc. Housing for the corporate entity refers to all costs necessary for physical facilities for its operations, and include rent, property taxes, mortgages payments, home insurance, maintenance costs, etc.

*Definition-4*: Leisure for the agent includes all costs for leisure and entertainment that are not necessary for day-to-day living, and include – entertainment costs, vacation time, etc.. Leisure for the corporate entity include all costs for entertainment and for employee leisure, such as corporate events, wellness programs, employee vacation time, etc..

*Definition-5*: For the agent, Intangibles includes all costs for developing or changing intangible property such as training expenses, costs for networking and special events, personal debt capacity, patent/trademark costs, costs for preparing proprietary data, etc.. For the firm, Intangibles includes all costs for developing or changing intangible property such as training expenses, corporate debt capacity, costs for networking and special events, patent/trademark costs, costs for preparing proprietary data, etc.

*Definition-6*: For the agent, Investment includes all traditional investment and savings activities and all forms of production and services activities because the agent is in effect investing time and Human Capital which provides returns in the form of salaries, and or royalties or fees or other remuneration. For the firm, Investment includes all traditional investment and savings activities and all forms of non-leisure production and services because the firm is in effect investing money, time, Human Capital, equipment and other resources.

*Definition-7*: For the agent, Taxes includes all traditional income taxes, capital gains taxes, property taxes and other taxes. For the Firm, Taxes includes all traditional income taxes, capital gains taxes, property taxes and other taxes. For the government, Taxes includes all tax incentives and tax abatements for traditional income taxes, capital gains taxes, property taxes and other taxes. Taxes can be positive, as in when wealth is allocated for payment of taxes or when government provides tax incentives/benefits; or negative as in when an agent avoids/defers payment of taxes; or when government eliminates existing tax incentives/benefits.

*Definition-8*: c, t, i, l, b and h are distinct single units of consumption, taxes, Investment, Leisure, Intangibles and Housing respectively.
$-\infty < v, y, x, z, s$ and $r < +\infty$ are the numbers of units of Total Wealth allocated to Consumption, Taxes, Investment, Leisure, Intangibles and Housing respectively, in the periodic wealth allocation process. These re-allocations are effected with single-period or multi-period contracts. Thus, the agent/investor can decide to forgo regular scheduled consumption (v can be negative); and the investor can decide to delay payment of taxes due, or implement tax reduction strategies (y can be negative); and can decide to withdraw funds from his/her investment securities account (x can be negative), and can decide to take a vacation, reduce regular work hours or spend more leisure time instead of working (z can be negative).

*Definition-9*: Markets are incomplete because there is intertemporal uncertainty (imperfect information about future states and preferences), there are contract enforcement costs, and the set of available contracts which can be used to transfer or re-allocate wealth across time is limited to those contracts that may match uncertain future states; and agents trade in both sequential spot markets and multi-period forward markets. T is a block of time that contains discrete units of time each of which is *t*.

***Theorem-1: The Consumption-Investment-Savings-Production Dichotomy Is Irrelevant Because All Four factors Are Elements Of One Decision Process And Are Insufficient For Defining Real World Situations; And Any Asset Pricing Model Based Solely On One Or All Of These Four Factors Is Inaccurate***.
***Proof***: Prior theoretical and empirical asset pricing studies have un-necessarily focused on the four approaches which are erroneously assumed to be different – the consumption-based, the savings-based (or consumption-savings-based), investment-based and production-based approaches to asset pricing. Basu (Oct. 2002). Garnier, Nishimura & Venditti (2007). Crossley & Low (Sept. 2006). See the discussion above on the feasibility conditions for the CSIP framework.

The UIWD is valid where the following conditions exist:
1) Individuals' Time and Knowledge have both monetary and non-monetary value. The individual's/investor's Time and Knowledge are part of his/her Total Wealth.
2) The capital markets are composed of firms that employ individuals based wholly or partly on their time, knowledge and effort.
3) For any individual and for any time interval, production, investment, savings and consumption can produce the same types and magnitudes of utility and wealth.
4) There are or may be opportunity costs for every allocation of an individual's resources/Total-Wealth to either consumption, savings, production or investment.
5) The individual's or investor's Marginal Rate of Intertemporal Substitution (MRIS) among any of production, investment, savings and consumption, changes in some proportion to his/her: i) Total Wealth, ii) Total Investable Wealth, iii) horizon. Here the MRIS is the rate at which the individual re-allocates one unit of Total Wealth (which includes Investible Wealth, Non-monetary Wealth and Monetary Wealth) among consumption, savings, investment or production.
6) Capital Markets are incomplete; and labor markets are also incomplete.
7) There is never pure equilibrium in capital markets or labor markets.
8) The household/investor derives the same types of utility/disutility from investments, products and traditional "goods", such that there is minimal distinction between "consumption", "investment" and "savings".
9) The household/investor derives the same types of utility/disutility from investment-products and traditional "goods" and "earnings" from work, such that there is minimal distinction among "consumption", "investment" and "savings" and "production".
10) The indifference curve between any pair of the investor's four allocation decision factors (consumption, savings, production and investment) is always downward-sloping.

All of the above-mentioned conditions exist simultaneously in many markets. □

7. The Elasticity of Intertemporal Substitution Is Inaccurate.
Other than the above, several authors have also noted the inaccuracy and inapplicability of the EIS. Guvenen (2006) attempted to reconcile two opposing views about the elasticity of intertemporal substitution (EIS) - empirical studies using aggregate consumption data typically find the EIS to be close to zero, whereas calibrated models designed to match growth and fluctuations facts typically require it to be close to one. Guvenent (2006) noted that this contradiction is resolved when two kinds of heterogeneity are acknowledged: one, the majority of households do not participate in stock markets; and secondly, that the EIS increases with wealth. When Guvenen (2006) introduced these two features into a standard real business cycle model, its was noted that limited participation creates substantial wealth inequality as in the U.S. data; and as such the properties of aggregates directly linked to wealth (e.g., investment and output) are mainly determined by the (high-EIS) stockholders; and since consumption is much more evenly distributed than is wealth, estimation from aggregate consumption uncovers the low EIS of the majority (i.e., the poor).

Havranek (2015) analyzed 2,735 (two thousand seven hundred and thirty five) estimates of the *Elasticity of Intertemporal Substitution* in consumption (EIS) that were derived in 169 (one hundred and sixty nine) published articles and found evidence of strong selective reporting wherein authors frequently discarded negative and insignificant estimates which in turn increased the mean estimate by about 0.5. Havranek (2015) noted that this "*Reporting Bia*s" was more dominant than the effects of empirical methods used (with the exception of the choice between micro and macro data). Most importantly, when Havranek (2015) corrected the mean for the *Reporting Bias*, for macro estimates the EIS obtained was zero even though the reported average *t-statistics* was two (2); but the corrected mean of micro estimates of the EIS for asset holders was around 0.3-0.4. Havranek (2015) concluded that estimated EIS that are greater than 0.8 are inconsistent with the bulk of the empirical evidence, and thus are wrong.

Wallenius (2011) considered two different skill accumulation technologies which are learning by doing and Ben-Porath type training. Wallenius (2011) noted that the effect of human capital accumulation in the form of learning-by-doing is to increase the labor supply elasticity estimate by a factor of 2.1 relative to the estimate that ignores human capital accumulation – all of which biases estimates of the EIS of Labor. Okubo (2008) used a model with nonseparable and nonhomothetic preferences to estimate the intertemporal elasticity of substitution (EIS), and found that while the assumption of homotheticity is strongly rejected, the estimated IES is positive and significant. Such empirical rejection of homotheticity is prime evidence that the EIS is wrong - because homotheticity is a major assumption underlying EIS. Lybbert & McPeak (2012) estimated risk aversion and intertemporal substitution as distinct preferences using data from Kenyan herders, and based on the assumption of existence of Epstein and Zin recursive preferences. Epstein and Zin [1989; Econometrica 57, 937–969], Lybbert & McPeak (2012) found that the assumption implicit in additive expected utility models that relative risk aversion (RRA) is the inverse of the elasticity of intertemporal substitution (EIS) is wrong. Lybbert & McPeak (2012) also stated that their RRA and EIS estimates are consistent with a preference for the early resolution of uncertainty, which is caused by the instrumental value of early resolution of uncertainty; and this same preference pattern is consistent with asset smoothing in response to a dynamic asset threshold. Such "early resolution" and "asset smoothing" trends/preferences render the EIS inaccurate because of the assumptions inherent in the definitions and calculation of EIS. Garcia, Renault & Semenov (2006) noted that although in the canonical CCAPM, the coefficient of relative risk aversion (RRA) is "constrained" to be the inverse of the EIS; for theoretical and empirical reasons the EIS and the RRA should be disentangled; and such disentangling may be achieved by replacing the future consumption stream not by a certainty equivalent of future utility, like in the recursive utility model of Epstein and Zin [1989. *Econometrica*, 57, 937–969], but by an exogenous reference level of consumption, which, in a recursive way, assesses the expected future consumption. Garcia, Renault & Semenov (2006) observations imply that those methods of calculating the EIS by direct or indirect reference to the RRA, are wrong.

Giulano & Turnovsky (2003) noted that the constant elasticity utility function implies that the intertemporal elasticity of substitution is the inverse of the coefficient of relative risk aversion, but empirical evidence suggests that this relationship may or may not hold, and thus studies of risk and growth should decouple these two parameters. Giulano & Turnovsky (2003) analyzed the "equilibrium" of a stochastically growing small open economy under general recursive preferences, and attempted to show that errors committed by using the constant elasticity utility function, even for small violations of the compatibility condition, can be substantial. The Giulano & Turnovsky (2003) results cast substantial doubt on the validity of both the EIS, and the EIS-RRA relationship.

Saltari & Ticchi (2007) analyzed the role of risk aversion and intertemporal substitution in a simple dynamic general equilibrium model of investment and savings, and found that risk aversion cannot by itself explain a negative relationship between aggregate investment and aggregate uncertainty, because the effect of increased uncertainty on investment also depends on the intertemporal elasticity of substitution. Saltari & Ticchi (2007) also noted that the relationship between aggregate investment and aggregate uncertainty is positive even if agents are very risk averse, as long as the elasticity of intertemporal substitution is low- but this statement is a contradiction to their other observations. Saltari & Ticchi (2007) stated that a negative investment–uncertainty relationship requires that the relative risk aversion and the elasticity of intertemporal substitution are both relatively high or both relatively low – but either condition contravenes the very definition of both RRA and EIS.

Lewis (1991) noted that empirical studies of the restrictions implied by the intertemporal capital asset pricing model across different asset markets have found conflicting evidence, and using data on foreign exchange, bonds, and equity returns, Lewis (1991) found that the tendency to reject the intertemporal consumption-based asset pricing relationship depends upon the inadequacy of an auxiliary assumption (that covariances of returns with consumption move in constant proportion over time), not necessarily the relationship itself. Stern (1997) noted that ecological economics is characterized by arguments concerning limits to substitution between inputs (energy, natural capital, etc., vs. manufactured capital, labor, etc.) in production and the implications these have for sustainability; and that various authors have also expressed concerns regarding limits to substitution in consumption, either between environmental assets and other goods or between basic needs commodities and other goods. Stern (1997) noted that the underlying theme is that individual commodities and other inputs have unique physical or other properties which make them poor substitutes, and that many authors have argued for irreversibilities in consumption behavior. Kim & Lee (2007) analyzed on-the-job human capital accumulation from the perspective of time invested for acquiring skills and learning by doing in an RBC model and shows that the inability to account for human capital accumulation leads to a substantial bias in conventional estimates of EIS. Kim & Lee (2007) stated that their main results are based on the standard intuition that the opportunity cost of time invested in acquiring human capital moves pro-cyclically, so that on-the-job time invested in acquiring human capital is "counter-cyclical"; and the true wage rate becomes less pro-cyclical, while production hours become more pro-cyclical than total hours at work.

Lee (2008) noted that estimates of EIS obtained from standard life-cycle models are subject to a downward bias because they neglect the life-cycle and demographic patterns of on-the-job human capital investment. Lee (2008) stated that there was statistically significant evidence that conventional estimates of EIS for total hours at work are biased downward by about 20% at the intensive margin; and the corresponding EIS estimates for production hours are biased downward even more, which provides an explanation for why output fluctuation is greater than hours/employment fluctuation over the business cycle. Taking into account the fact that part of a worker's time at work goes to acquiring human capital in addition to his main task of producing goods, Lee (2008) attempted to extend the standard life-cycle model to include time spent on investing in on-the-job human capital and proposed a new framework for identifying the EIS.

***Theorem 2: The Elasticity Of Intertemporal Substitution (EIS) Is Inaccurate***.
***Proof*:** Many studies have empirically and theoretically derived very different estimates for the Elasticity of Intertemporal Substitution (EIS). Giuliano & Turnovsky (2003). Crossley & Low (Sept. 2006). The EIS is deficient because it addresses only substitution between only two factors/goods in only one domain (typically the consumption domain, or production domain or investment domain). The EIS does not address the UIWD which refers to re-allocation of wealth among consumption, production, investment and or savings. The EIS does not account for the finiteness of the factors being substituted – ie. a human being can work for only a finite number of hours in each day/month/year; a household has a finite amount of wealth that it can spend (even when borrowing and leasing are considered); etc..

Even if the EIS is meaningful, the EIS in either consumption-savings domain, or investment domain or production domain is never constant because of the following reasons. Labor markets are incomplete and constantly evolving, and in most circumstances, UIWD prevails. EIS does not account for different perceptions of time by different people/groups; and the different values of different types of "time" (ie. leisure time; work time; family time) to different people/groups – this is a critical factor in intertemporal analysis. EIS addresses only changes in two periods, which is, or can be very misleading because many decisions and preferences substantially multi-period in nature (ie. involve more than three time periods). EIS is defined completely without reference to Investible Wealth or Total Wealth. This is error, because there are many behavioral biases (Anchoring Effects and Framing-Effects) that are intentional (advertising from brokerage firms) or un-intentional (discussion with spouse/partners) and that observed or un-observed, in the wealth re-allocation decision process. Most of these Framing Effects are based on Total Wealth, or Investible Wealth, or Disposable Wealth or Future Wealth. Furthermore, consumption, Savings, Investment or Production is best defined with reference to some form of wealth or Total Wealth.

The EIS does not consider Regret at both the individual and group levels.

Giuliano & Turnovsky (2003) notes that the intertemporal elasticity of substitution, emphasized by Hall (1978, 1988), and Mankiw, Rotemberg & Summers (1985) and others, focuses on intertemporal preferences and is well defined in the absence of risk. This is error because the investor's wealth re-allocation decision processes almost always involves some analysis of risk which is manifested in part by the actual re-allocation process. Giuliano & Turnovsky (2003) notes that a natural definition of the EIS is in terms of the percentage change in intertemporal consumption in response to a given percentage change in the intertemporal price. Hence, EIS is erroneously defined primarily in terms of consumption and prices, which by themselves are insufficient to fully capture the investor's preferences,

EIS does not account for the fact that the average investor's labor income is subject to various shocks; and that the investor is able to observe most of the components of his/her labor income in the short run (1-15 days). EIS does not account for Anchoring Effects and Framing Effects. EIS does not account for the fact that the average investor's propensity to substitute is partly based on his/her Total Wealth, wealth available for re-allocation, and perceived risk of both opportunity costs and returns from abstinence.

EIS may be accurate only if the permanent-income hypothesis (PIH) (introduced by Friedman (1957)) is correct. The PIH states that consumption is equal to the annuity value of ''total wealth'' calculated as the sum of the discounted expected value of future income (discounted using the risk-free rate), plus the agent's "Human Wealth" plus financial wealth (cumulative savings). Contrary to the existing literature and given the analysis herein and above (including the *Unified Intertemporal Wealth-Allocation Decision* Framework; and the invalidity of the *CSIP Framework*), the individual agent's optimal consumption rule is not governed by the PIH. The conditions under which the PIH rule is feasible are un-realistic, and the PIH erroneously implies that changes in an individual agent's consumption are not predictable (Hall, 1978). The PIH is based on the erroneous assumptions that: **i)** utility is quadratic utility; and **ii)** there is no precautionary savings by the average individual, and **iii)** the individual's future labor income is riskless, and **iv)** the subjective discount rate and the prevailing interest rate are equal. Guvenen (2006) and Havranek (2015) also noted that there is conflicting evidence about the validity of EIS. □

8. The Relationships Among The Factors.

***Theorem-3: For Any Time Interval Or Successive Time Intervals, each of c, t, i, l, b and h is non-monotonic, And The relationship between each of c, t, i, l, b and h on one hand and Total Wealth on the other hand, Is Also Non-Monotonic, where Total Wealth Is Finite, Worker Effort Is Rewarded With Monetary Benefits And Utility/Disutility provided By Consumption/Taxes/Investment/Leisure/Intangibles/Housing Is Non-Monotonic.***
*Proof*: For any time interval or any series of successive time intervals, the relationship between total Wealth and each of ***c, t, i, l, b and h*** is not monotonic because there may be frictions, surprises and other considerations that may cause the relationship to change, such as new knowledge; fairness; pressure from spouse/partner; willingness to defer gains/losses; need for leisure time; involuntary changes in work conditions; temporary or permanent layoffs; advertising; change of job; etc.. Similarly the relationships between Total Wealth and each of ***c, t, i, l, b and h*** could change. The relationships between Total Wealth and on the other hand, each of i and l could change due to new knowledge; pressure from spouse/partner; need for leisure time; involuntary changes in work conditions; temporary or permanent layoffs; advertising; change of job; etc.. Therefore, $di/dW$, $dp/dW$, $dt/dW$, $dl/dW$, $dh/dW$ and $dc/dW$ are all non-monotonic. □

***Theorem-4: For Any Time Interval Or Successive Time Intervals, each of c, t, i, l, b and h is non-additive, And The Relationship Between Total Wealth and Each Of c, t, i, l, b and h Is Also Non-Additive, where Total Wealth Is Finite, Worker Effort Is Rewarded With Monetary Benefits And Utility/Disutility provided By Consumption/Taxes/Investment/Leisure/Intangibles/Housing Is Non-Monotonic.***
*Proof*: Each of ***c, t, i, l, b and h*** is non-additive in any time interval – for example, where c(.) is a consumption function, $c(x) + c(y) \neq c(x+y)$. First, allocations of wealth to ***c, t, i, l, b and h*** are done primarily at the beginning of the subject period (and also during the subject period), and any additional allocations require either reduction of allocations to the other five factors or a change in Total Wealth, which in turn, causes changes in Anchoring, Framing and preferences, all of which changes the utility/disutility of the sum of the added allocations {c(x+y)}. The second reasons is that the average investor derives different amounts of utility from, and assigns different risk profiles for each of ***c, t, i, l, b and h***, and such utility are dynamic. Thus, when c(x) is added to c(y) the relative risk of ***c, t, i, l, b and h*** is very likely to change such that the sum will not be c(x+y). It also follows automatically that the relationship between Total Wealth and each of ***c, t, i, l, b and h*** is also non-additive. □

***Theorem-5: For Any Time Interval Or Successive Time Intervals, The relationship between Total Wealth, and each of c, t, i, l, b and h is recursive, where Total Wealth Is Finite, Worker Effort Is Rewarded With Monetary Benefits, And the Utility/Disutility provided By Consumption/Taxes/Investment/Leisure/Intangibles/Housing Is Non-Monotonic.***
***Proof***: $-\infty < v, y, x, z, s, r < +\infty$ are the numbers of units of Total Wealth allocated to Consumption, Taxes, Investment, Leisure, Intangibles and Housing respectively, in the periodic wealth allocation process. Because of UIWD, in every sub-period *t*, the amount of Total Wealth allocated to a factor is a direct function of:

a) amounts of wealth allocated to all six factors in the prior period – the utility/disutility gained from such prior allocations shape decisions about future allocations and expected Regret.
b) Amounts of Wealth allocated to the other five factors in the current period (three of $c_{(t+1)}$, $t_{(t+1)}$, $i_{(t+1)}$, $l_{(t+1)}$, $b_{(t+1)}$, $h_{(t+1)}$) – which also affected re-allocations in the current period.
c) The portion of Total Wealth that is available for allocation in the current period ($w_{(t+1)}$).
d) Total Wealth ($W_T$)
Therefore, c, t, i, l, b and h are recursive, and:

$x_{(t+1)} = f(x_t, v_t, y_t, z_t, r_t, s_t, w_{(t+1)}, v_{(t+1)}, y_{(t+1)}, z_{(t+1)}, r_{(t+1)}, s_{(t+1)}, W_T)$
$v_{(t+1)} = f(x_t, v_t, y_t, z_t, r_t, s_t, w_{(t+1)}, x_{(t+1)}, y_{(t+1)}, z_{(t+1)}, r_{(t+1)}, s_{(t+1)}, W_T)$
$y_{(t+1)} = f(x_t, v_t, y_t, z_t, r_t, s_t, w_{(t+1)}, v_{(t+1)}, x_{(t+1)}, z_{(t+1)}, r_{(t+1)}, s_{(t+1)}, W_T)$
$z_{(t+1)} = f(x_t, v_t, y_t, z_t, r_t, s_t, w_{(t+1)}, v_{(t+1)}, y_{(t+1)}, i_{(t+1)}, r_{(t+1)}, s_{(t+1)}, W_T)$
$r_{(t+1)} = f(x_t, v_t, y_t, z_t, r_t, s_t, w_{(t+1)}, v_{(t+1)}, y_{(t+1)}, i_{(t+1)}, x_{(t+1)}, s_{(t+1)}, W_T)$
$s_{(t+1)} = f(x_t, v_t, y_t, z_t, r_t, s_t, w_{(t+1)}, v_{(t+1)}, y_{(t+1)}, i_{(t+1)}, r_{(t+1)}, x_{(t+1)}, W_T)$

Also, the relationship between Total Wealth (W) and each of c, t, i, l, b and h is recursive in any time interval or a series of successive time intervals – thus where a, b, d, e, j, k are variables with defined formulas:
$dx_{(t+1)}/dW_{(t+1)} = a*f\{dx_t/dW_t\}$ because for most investors, investment returns in the first period and the change in Total Wealth has a direct effect on the amount of wealth that is allocated to Investments in the second period.
$dv_{(t+1)}/dW_{(t+1)} = b*f\{dx_t/dW_t\}$ because for most investors, the utility/disutility gained from consumption in the first period and the change in Total Wealth have a direct effect on the amount of wealth that is allocated to Consumption in the second period and subsequent periods.
$dy_{(t+1)}/dW_{(t+1)} = d*\{dy_t/dW_t\}$ because for most investors, the amount of wealth allocated to Taxes and the associated returns in the first period and the change in Total Wealth both have a direct effect on the amount of wealth that is allocated to Taxes in the second and subsequent periods.
$dz_{(t+1)}/dW_{(t+1)} = e*\{dz_t/dW_t\}$ because for most investors, the amount of wealth allocated to Leisure and the associated returns in the first period and the change in Total Wealth both have a direct effect on the amount of wealth that is allocated to Leisure in the second period and subsequent time periods.   □

***Theorem-6: a) The Marginal Rate Of Intertemporal Joint Substitution measures the propensity for an investor/household to substitute units of Consumption, Taxes, Investment, Leisure, Intangibles and Housing (factors), where Total Wealth is finite and limited in each time interval, and the effects of such substitution may increase or decrease the other five factors; b) the Marginal Rate Of Inter-temporal Joint Substitution (MRIJS) measures the investor's propensity to substitute/re-allocate one unit of Total Wealth from any of five factors (c, l, t, i, b or h) to a sixth factor.***
*Proof*: $-\infty < v, y, x, z, s, r < +\infty$ are the numbers of units of Total Wealth allocated to Consumption, Taxes, Investment, Leisure, Intangibles and Housing respectively, in the periodic wealth allocation process. W(x) is a wealth function; and $w_t$ denotes the Periodic Total Wealth to be allocated at the beginning of the budget period. The Utility gained by the household from such wealth is a function defined as:
U(c,t,I, l,b,h) = F{W(c,t,I, l,b,h)}

$w_t = vc + yt + xi + zl + sb + rh$
$v = (W_t - xi_t - sb_t - l_tz_t - yt_t - rh_t)$
$x = (W_t - vc_t - sb_t - lz_t - yt_t - rh_t)$
$y = (W_t - vc_t - i_tx - sb_t - l_tz - rh_t)$
$z = (W_t - vc_t - sb_t - i_tx - yt - rh_t)$
$s = (W_t - vc_t - i_tx - l_tz - yt_t - rh_t)$
$r = (W_t - vc_t - sb_t - i_tx - yt_t - l_tz)$

Let:
$C\check{} =$ the rate of substitution of Consumption with respect to the other five factors – that is, the average change in Consumption, as a result of simultaneous changes in the other five factors.
$I\check{} =$ the rate of substitution of Investment with respect to the other five factors – that is, the average change in

Investment, as a result of simultaneous changes in the other five factors.
T˜ = the rate of substitution of Taxes with respect to the other five factors – that is, the average change in Taxes, as a result of simultaneous changes in the other five factors.
L˜ = the rate of substitution of Leisure with respect to the other five factors – that is, the average change in Leisure expenditures, as a result of simultaneous changes in the other five factors.
B˜ = the rate of substitution of Intangibles with respect to the other five factors – that is, the average change in expenditure on Intangibles, as a result of simultaneous changes in the other five factors.
H˜ = be the rate of substitution of Housing with respect to the other five factors – that is, the average change in Housing expenditures as a result of simultaneous changes in the other five factors.

C˜, I˜, T˜, L˜, B˜ and H˜ are measured over one or more periods and thus, there is some averaging.
Then:
$\mathbb{C}$˜ = $\partial v/\partial W_t - \partial v/\partial x_t - \partial v/\partial y_t - \partial v/\partial z_t - \partial v/\partial s_t - \partial v/\partial s_t$
L˜ = $\partial z/\partial W_t - \partial z/\partial x_t - \partial z/\partial y_t - \partial z/\partial v_t - \partial z/\partial s_t - \partial z/\partial r_t$
T˜ = $\partial y/\partial W_t - \partial y/\partial x - \partial y/\partial v_t - \partial y/\partial z_t - \partial y/\partial s_t - \partial y/\partial r_t$
I˜ = $\partial x/\partial W_t - \partial x/\partial v_t - \partial x/\partial y_t - \partial x/\partial z_t - \partial x/\partial s_t - \partial x/\partial r_t$
B˜ = $\partial s/\partial W_t - \partial s/\partial v_t - \partial s/\partial y_t - \partial s/\partial z_t - \partial s/\partial x_t - \partial s/\partial r_t$
H˜ = $\partial x/\partial W_t - \partial r/\partial v_t - \partial r/\partial y_t - \partial r/\partial z_t - \partial r/\partial s_t - \partial r/\partial x_t$

C˜, L˜, T˜, I˜, B˜, H˜ are referred to as the Marginal Rates Of Substitution.

MRIJS is the *Marginal Rate Of Inter-temporal Joint Substitution*. MRIJS is the investor's propensity to substitute/re-allocate one unit of Total Wealth from any of five factors (c, l, t, i, b or h) to a sixth factor. MRIJS ε (0, 1).

MRIJS = exp[Min{0, -(C˜ + L˜ + T˜ + I˜ + B˜ + H˜)}]

Thus, MRIJS is a measure of both a person's ability-to-repay and willingness-to-repay an obligation. The greater a person's total wealth, the greater his/her ability to reallocate such wealth among the six factors. The MRIJS implicitly incorporates Regret, because the person's reallocation of wealth among the six factors implicitly includes a regret minimization process.

▫

9. Other Recent Research On Asset Pricing.
Given the foregoing analysis, the theories and models introduced and or discussed in the following articles, are either moot or inaccurate: Meghir & Weber (1996); Ray & Robson (2012); Chen & Epstein (2002); Ozsoylev & Walden (2011); Hugonnier (2012); Adam & Marcet (2011); Epstein & Zin (1989); Duffie & Strulovici (2012); Bossaerts, Plott & Zame (2007); Alvarez & Jermann (2005); Alvarez & Jermann (2000).

10. Conclusion.
In most markets, investors' preferences diverge substantially, all existing asset pricing models are inaccurate because the underlying assumptions are not realistic. Obviously, this has important ramifications for asset management and capital budgeting. The *Elasticity of Intertemporal Substitution* is very inaccurate especially where markets are incomplete and investors' preferences are both dynamic and multi-faceted. The focus on consumption and price as the definition of investors' preferences and constraints is very limited and misleading – the investors' decision problem is much broader in scope.
Given the problems and inaccuracies inherent in the Consumption-Savings-Investment-Production dichotomy, the more unified asset-pricing approach introduced herein (UIWD) is more likely to result in better policy decisions. The MRIJS is distribution-free, does not require use of any specific utility functions, implicitly accounts for risk (by multi-faceted wealth allocation) and provides a more unified and accurate indication/analysis of the average investor's wealth allocation decisions. MRIJS and theorems 4 & 5 & 6 and the theories introduced herein are also testable hypothesis, which can become the foundation for further detailed models that are better able to reflect reality and economic transactions.